\documentclass[11pt,a4paper,3p, leqno]{iranote}
\usepackage{theorem}
\usepackage{colonequals}
\usepackage{microtype,ifxetex}
\usepackage{todonotes,url,tcolorbox,newalg,afterpage,array,caption}
\usepackage[colorlinks=false,allbordercolors={1 1 1}]{hyperref}
\usepackage{amsmath,amsfonts,amssymb}
\usepackage[vlined,linesnumbered]{algorithm2e}
\usepackage[capitalize]{cleveref}
\usepackage{doi}
\usepackage{hyperref}
\newcommand{\var}{\mathrm{var}}
\newcommand{\varall}[1][\Psi]{V^{\forall}_{#1}}
\newcommand{\varex}[1][\Psi]{V^{\exists}_{#1}}
\newcommand{\dep}[1][\Psi]{\mathrm{dep}_{#1}}
\newcommand{\upy}{\mathrel{\vdash_{\!\!\!\!\raisebox{-1pt}{\scriptsize $1\forall$}}}}
\newcommand{\upd}{\mathrel{\vDash_{\!\!\!\!\raisebox{-2pt}{\scriptsize $1\forall$}}}}
\newcommand{\prefix}{\Pi}

\DeclareMathOperator{\abs}{abs}
\DeclareMathOperator{\OV}{OV}
\DeclareMathOperator{\OC}{OC}
\DeclareMathOperator{\OR}{OR}
\newcommand{\UR}[1][\prefix]{\mathrm{UR}_{#1}}
\DeclareMathOperator{\UP}{UP}
\newcommand{\assign}{\mathcal{A}}

\newcommand{\dqior}{DQIOR}

\newcommand{\dqratPlus}{DQRAT\textsuperscript{+}}

\ifxetex
    \newcommand{\conflict}{\,\downzigzagarrow}
\else
    \usepackage{stmaryrd}
    \newcommand{\conflict}{\lightning}
\fi
\makeatletter
\newcommand\xleftrightarrow[2][]{%
  \ext@arrow 9999{\longleftrightarrowfill@}{#1}{#2}}
\newcommand\longleftrightarrowfill@{%
  \arrowfill@\leftarrow\relbar\rightarrow}
\makeatother

\begin{document}
\title{Using Unit Propagation with Universal Reduction in DQBF Preprocessing}
\author{Ralf Wimmer${}^{1,2}$, Ming-Yi Hu\textsuperscript{2}}
\address{\textsuperscript{1}Altair Engineering GmbH, B\"oblingen, Germany}
\address{\textsuperscript{2}Albert-Ludwigs-Universit\"at Freiburg, Freiburg im Breisgau, Germany\\
  \texttt{\{wimmer, hum\}@informatik.uni-freiburg.de}}
\date{\today}
\begin{abstract}
    Several effective preprocessing techniques for Boolean formulas with and
    without quantifiers use unit propagation to simplify the formula. Among these
    techniques are vivification, unit propagation look\-ahead (UPLA), and the identification
    of redundant clauses as so-called quantified resolution asymmetric tautologies (QRAT).
    For quantified Boolean formulas (QBFs), these techniques have
    been extended to allow the application of universal reduction during unit
    propagation, which makes the techniques more effective.

    In this paper, we prove that the generalization of QBF to
    dependency quantified Boolean formulas (DQBFs) also allows
    the application of universal reduction during these preprocessing
    techniques.
\end{abstract}
\maketitle

\section{Introduction}
\label{sec:introduction}

\noindent Algorithms for solving the propositional satisfiability problem
(SAT) have seen enormous improvements during the last three decades~\cite{handbook_of_sat}
and have turned into a standard means for solving NP-complete problems not only in
academia, but also in industry.

Motivated by this success, research has turned its attention to the harder
problem of solving quantified propositional formulas -- first to formulas with a
linear quantifier prefix where the exististential variables are linearly ordered \wrt
the sets of universal variables upon which they depend. This class of formulas
is known as QBFs, and deciding satisfiability is known to be PSPACE-complete~\cite{MeyerS73}.

For roughly one decade, researchers have also investigated a more general form
of propositional formulas with quantifiers, so-called \emph{dependency quantified
Boolean formulas} (DQBFs)~\cite{Henkin61}, where each existential variable can depend on an arbitrary
subset of the universal variables. This flexibility makes the decision problem
NEXPTIME-complete~\cite{PetersonRA01}. However, the expressive power of DQBFs is necessary for expressing a number
of relevant problems in a succinct way. Examples include the verification of incomplete
circuits~\cite{gitina-et-al-iccd-2013,wimmer-et-al-synthesisBook-2017}, synthesis problems~\cite{FaymonvilleFRT17},
solving games with incomplete information~\cite{PetersonRA01}, and many more~\cite{ChenHLT22}.

While for SAT problems, search-based tools seem to be the most successful algorithms,
DQBF tools successfully implement a wide rage of different ideas: DPLL-style search~\cite{FrohlichKB12},
instantiation (iDQ)~\cite{FrohlichKBV14},
quantifier-elimination using and-inverter graphs (HQS)~\cite{gitina-et-al-date-2015,geernst-et-al-tcs-2022} and
reduced ordered binary decision diagrams (DQBDD)~\cite{SicS21}, abstraction (dCAQE)~\cite{TentrupR19},
or the extraction of uniquely defined Skolem functions plus
counterexample-guided synthesis (PEDANT)~\cite{ReichlS22}.

Part of the success of these tools is due to intensive preprocessing of the
formulas to simplify them before the actual solving process, \eg by eliminating
variables that can be eliminated at low cost, by deleting redundant clauses,
by strengthening (\ie shortening) clauses and many more. The state of the art
in DQBF preprocessing is described in \cite{wimmer-et-al-jsat-2019} and
implemented in the tool HQSpre.

Still, some of the applied techniques mimic the corresponding SAT-based techniques
without a real generalization to the possibilies that DQBF offers: \emph{Universal reduction}
removes universal literals from clauses on which none of the existential literals of
the clause depends, yielding an equivalent formula. Applying universal reduction during
unit propagation, which removes all unit clauses from the formula, is standard in
(D)QBF preprocessing. However, there are further techniques that make use of unit
propagation: Vivification~\cite{PietteHS08}
strives to shorten or remove clauses; unit propagation-lookahead (UPLA)~\cite{Berre01}
aims at finding implied or equivalent literals; (D)QRAT~\cite{HeuleSB14,Blinkhorn20} identifies
redundant clauses that can removed without changing satisfiability of the formula.
For DQBF, all of them currently use unit propagation without universal reduction.
Universal reduction would make these techniques more effective as it allows to find
more implications or conflicts. Applying it blindly, however, makes these techniques
unsound.

The results in this paper show for DQBF how to combine universal reduction with
vivification, UPLA, and DQRAT. We provide detailed proofs of correctness.
Thereby we generalize similar results by Lonsing and Egly~\cite{LonsingE18,LonsingE19,Lonsing19}
from QBF to DQBF. While the results are the natural generalization of the corresponding
results for QBF, the proofs for DQBF differ considerably from those for QBF as the latter exploit
the linearity of the quantifier prefix and the fact that assignment trees can be used
as a model for a satisfiable QBF. For DQBF, the proofs have to reason using
Skolem functions instead. \dqratPlus{} as defined in this paper is also
a generalization of DQRAT~\cite{Blinkhorn20}.

\section{Foundations}
\label{sec:foundations}

\noindent Let $\bools$ denote the set $\{0,1\}$ of truth values. For a finite set
$V$ of Boolean variables, $\assign(V)$ denotes the set of variable assignments
of $V$, \ie $\assign(V)=\{\lambda\,|\,\lambda:V\to\bools\}$.
Dependency quantified Boolean formulas are obtained by prefixing Boolean
formulas with so-called Henkin quantifiers~\cite{Henkin61}.

\begin{definition}[Syntax of DQBF]
  \label{def:dqbf_syntax}
  Let $V = \{x_1,\ldots,x_n,y_1,\ldots,y_m\}$ be a finite set of Boolean variables.
  A \emph{dependency quantified Boolean formula (DQBF)} $\Psi$ over $V$ has the form
  \[
    \Psi\colonequals\forall x_1\ldots\forall x_n\exists y_1(D_{y_1})\ldots\exists y_m(D_{y_m}):\varphi
  \]
  where $D_{y_i}\subseteq\{x_1,\ldots,x_n\}$ is the \emph{dependency set} of $y_i$
  for $i=1,\ldots,m$, and $\varphi$ is a quantifier-free Boolean formula over $V$.
\end{definition}

Because all dependencies of variables are explicitly specified (in contrast to QBF where
the dependencies are implicitly given by the order of the variables), the order of variables
in the prefix does not matter. Therefore we can consider the \emph{prefix} $\forall x_1\ldots\exists y_m(D_{y_m})$
as a set $\prefix$ and write $\prefix\setminus\{v\}$ for a variable $v$.
If $v$ is universal, this means that we remove $\forall v$
from the prefix and also $v$ from the dependency sets of all existential variables in which $v$
appears. If $v$ is existential, we remove $\exists v$ together with its dependency set from the
quantifier prefix. Other set operations have the intuitive meaning.

We denote by $\varex$ the set of existential and by $\varall$ the set of universal variables
of $\Psi$.

In this paper, we assume that the \emph{matrix} $\varphi$ of $\Psi$ is in conjunctive normalform
(CNF): A \emph{literal} is either
a variable $v$ or its negation $\neg v$. For a literal $\ell$ we write $\var(\ell)$ to get its
corresponding variable. A \emph{clause} is a disjunction of literals. As usual, we treat clauses
also as sets of literals because their order does not matter and duplicate literals can be removed.
A formula is in \emph{conjunctive normalform} if it is the conjunction of clauses. Similarly,
we treat a formula in CNF as a set of clauses. We call a clause $C$ \emph{compabible} with a
DQBF $\Psi$ if $C$ only contains variables of $\Psi$.

\begin{definition}[Semantics of DQBF]
    \label{def:dqbf_semantics}
    Let
    \[
        \Psi \colonequals \forall x_1\ldots\forall x_n\exists y_1(D_{y_1})\ldots\exists y_m(D_{y_m}): \varphi
    \]
    be a DQBF. A \emph{Skolem function} $(s_y)_{y\in\varex}$ for $\Psi$ maps every existential variable $y_i$ to a
    function $s_{y_i}: \assign(D_{y_i})\to \{0,1\}$ such that replacing every existential variable
    $y_i$ of $\Psi$ by (a Boolean expression for) $s_{y_i}$ turns $\varphi$ into a tautology.

    We call $\Psi$ \emph{satisfiable} if a Skolem function for $\Psi$ exists, otherwise
    \emph{unsatisfiable}.
\end{definition}

\noindent To simplify notations, we generalize the dependencies of a DQBF $\Psi$ from existential to all variables as follows:
\begin{align*}
    \dep(v) &\colonequals \begin{cases}
        \{v\} & \text{if $v$ is universal,} \\
        D_v   & \text{if $v$ is existential.}
    \end{cases} \\
\intertext{Additionally, we set for a literal $\ell$ and a clause $C$:}
    \dep(\ell) &\colonequals \dep\bigl(\var(\ell)\bigr), \\
    \dep(C)    &\colonequals \bigcup_{\ell\in C}\dep\bigl(\ell\bigr).
\end{align*}

\begin{definition}[Implication, equivalence, equi-satisfiability]
    Let $\Psi_1$ and $\Psi_2$ be two DQBFs over the same sets of existential and universal variables.
    $\Psi_1$ \emph{implies} $\Psi_2$ (written $\Psi_1\vDash\Psi_2$)
    if every Skolem function for $\Psi_1$ is also a Skolem function for $\Psi_2$.

    $\Psi_1$ and $\Psi_2$ are \emph{equivalent} (written $\Psi_1\equiv\Psi_2$)
    if they have the same Skolem functions, \ie if
    $\Psi_1\vDash\Psi_2$ and $\Psi_2\vDash\Psi_1$.

    $\Psi_1$ and $\Psi_2$ are \emph{equi-satisfiable} (written $\Psi_1\approx\Psi_2$)
    if either both $\Psi_1$ and $\Psi_2$ are satisfiable or both are unsatisfiable.
\end{definition}

\begin{definition}[Universal Reduction]
    \label{def:universal_reduction}
    Let $\Psi \colonequals\prefix:\varphi$ be a DQBF as defined above and $C$ a non-tautological clause, compatible
    with $\Psi$. \emph{Universal reduction} produces the clause
    \[
        \UR(C) \colonequals
        C\setminus \bigl\{\ell\in C\,\big|\,
            \var(\ell)\in\varall\land \nexists k\in C:(\var(k)\in\varex\land \var(\ell)\in D_{\var(k)})\bigr\}.
    \]
    Universal reduction applied to $\Psi$ yields
    $\UR(\Psi) \colonequals\prefix:\bigl\{\UR(C)\,\big|\,C\in\varphi\bigr\}$.
\end{definition}

\begin{lemma}
    \label{lemma:universal_reduction}
    Let $\Psi$ be a DQBF. Then $\Psi\equiv\UR(\Psi)$.
\end{lemma}
\begin{proof}
    See \cite{wimmer-et-al-jsat-2019}.\qed
\end{proof}

\begin{definition}[Unit Propagation]
    \label{def:unit_propagation}
    Let $\Psi\colonequals\prefix:\varphi$ be a DQBF and
    $U\colonequals \bigl\{\ell\,\big|\, \{\ell\}\in\varphi\land\var(\ell)\in\varex\bigr\}$ be the set of
    existential unit clauses of $\Psi$. \emph{Unit propagation} applied to $\Psi$
    yields the DQBF:
    \begin{align*}
    \UP^1(\Psi) & \colonequals
        \UR\Bigl(\prefix\setminus\{\var(\ell)\,|\,\ell\in U\}:
        \bigl\{C\setminus\{\neg\ell\,|\,\ell\in U\}\,\big|\,C\in\varphi\land C\cap U=\emptyset\bigr\}
        \Bigr). \\
    \intertext{
    Since unit propagation can yield new unit clauses, we iterate its application:
    }
        \UP^{n+1}(\Psi) &\colonequals \UP(\UP^n(\Psi)). \\
    \intertext{
        Iterating until the formula does not change anymore yields:
    }
        \UP(\Psi) &\colonequals \lim_{n\to\infty}\UP^n(\Psi).
    \end{align*}
    If $\emptyset\in\UP(\Psi)$, unit propagation yields a conflict and
    we write $\Psi\upy\conflict$. Otherwise,
    if $U$ is the set of all unit literals processed in all rounds of
    unit propagation, we write $\UP(\Psi)\upy U$.
\end{definition}
The iteration terminates after a finite number of steps when the formula
is not modified anymore. Note that $U$ only contains existential literals.
Universal unit clauses are reduced to empty clauses by universal reduction,
yielding a conflict instead.

\begin{lemma}
    \label{lemma:unit_propagation}
    Let $\Psi\colonequals\prefix:\varphi$ be a DQBF.

    If $\Psi\upy\conflict$, then $\Psi$ is unsatisfiable.

    If $\Psi\upy U$, then $\Psi\equiv\prefix:\varphi\land\bigwedge_{\ell\in U}\ell$.
\end{lemma}

\begin{proof}
    See \cite{wimmer-et-al-jsat-2019}.\qed
\end{proof}

Unit propagation for DQBF not only processes the unit clauses of a formula,
but also applies universal reduction to remove redundant universal literals.
By this, in general, more unit literals can be found and processed.

\section{Preprocessing Techniques}

\noindent Our goal is to apply unit propagation with universal reduction as
defined in Def.~\ref{def:unit_propagation} as part of preprocessing techniques
like vivification~\cite{PietteHS08}, unit propagation look-ahead~\cite{Berre01}, and dependency
quantified resolution asymmetric tautologies (DQRAT)~\cite{Blinkhorn20}. Doing so na\"\i{}vely,
make the techniques unsound.

\subsection{Formula Abstraction}
\label{sec:abstraction}

To make them sound, we first have to turn some
universal variables of the formula into existential variables without dependencies.
We call this step \emph{abstraction}:

\begin{definition}[Abstraction]
    \label{def:abstraction}
    Let $\Psi \colonequals \prefix:\varphi$ be a DQBF over variables $V$ as defined
    above and $V'\subseteq\varall$.
    The \emph{abstraction} $\abs(\Psi, V')$ of $\Psi$ \wrt $V'$ is the DQBF
    \[
        \bigl(\prefix\setminus V'\bigr) \dcup \bigl\{ \exists v(\emptyset)\,\big|\, v\in V'\bigr\}: \varphi\,.
    \]
\end{definition}
That means the abstraction of $\Psi$ \wrt $V'$ turns all variables in $V'$ into existential
variables without dependencies. The matrix of $\Psi$ is left unchanged.

The case where only the matrix of a DQBF is taken into account and the prefix is
ignored, is a special case of abstraction when $V' = \varall$. This turns the
DQBF into a SAT problem.

\begin{lemma}
    \label{lemma:abs_sat}
    Let $\Psi$ be a DQBF over variables
    $V$ as defined above and $V'\subseteq\varall$.
    If $\Psi$ is satisfiable, then $\abs(\Psi,V')$ is satisfiable as well.
\end{lemma}
\begin{proof}
    Assume that $\Psi \colonequals \prefix:\varphi$ is satisfiable.
    Then there is a Skolem function $(s_y)_{y\in\varex}$
    for $\Psi$. We construct a Skolem function for $\abs(\Psi,V')$. Let
    $\lambda:V'\to\bools$ be an arbitrary assignment of the variables
    in $V'$. For existential variables $v\in\varex[\abs(\Psi,V')] = V'\dcup\varex$, we set
    \[
        s'_v\colonequals
        \begin{cases}
            \lambda(v) & \text{if $v\in V'$,} \\
            s_{v|\lambda} & \text{if $v\in\varex$,}
        \end{cases}
    \]
    where $s_{v|\lambda}$ means that we take $s_v$ and replace
    every universal variable $x\in D_v$ for which $\lambda$ is defined by $\lambda(x)$.

    By that we have defined functions $(s'_v)_{v\in\varex\cup V'}$. They satisfy
    the dependencies specified in the prefix of $\abs(\Psi,V')$:
    If $v\in V'$, $s'_v$ is a constant function with value $\lambda(v)$;
    if $v\in \varex$, all variables $x\in D_v\cap V'$ have been replaced by
    the constant $\lambda(x)$, yielding a function $s'_v:\assign(D_v\setminus V')\to\bools$.

    Finally, we have to show that the defined functions are actually Skolem
    functions for $\abs(\Psi,V')$. Take an arbitrary assignment $\mu$ of the
    universal variables in $\abs(\Psi,V')$ and set $\kappa = \lambda\cup\mu$, \ie
    $\kappa$ is defined for all universal variables of $\Psi$; it coincides with
    $\lambda$ for all universal variables in $V'$ and with $\mu$ for all other universal
    variables. It follows that $s_y(\kappa) = s_y'(\mu)$ for all $y\in\varex$. Since
    replacing all universal variables $x\in\varall$ by $\kappa(x)$ and all existential variables
    $y\in\varex$ by $s_y(\kappa)$ satisfies $\varphi$, the same holds for $\abs(\Psi,V')$ and
    the functions $s'_v$ for $v\in\varex\cup V'$. Consequently, the constructed
    functions $(s'_z)_{z\in\varex\cup V'}$ are Skolem functions for $\abs(\Psi,V')$.\qed
\end{proof}
A consequence of \cref{lemma:abs_sat} is that if $\abs(\Psi,V')$ is unsatisfiable,
$\Psi$ is unsatisfiable as well.

\begin{lemma}
    \label{lemma:abs_equiv}
    Let $\Psi_1 \colonequals \prefix:\varphi_1$ and
    $\Psi_2 \colonequals \prefix:\varphi_2$ be two DQBFs over variables
    $V$ with identical prefixes, but possibly different matrices $\varphi_1$ and
    $\varphi_2$. Let $V'\subseteq\varall[\prefix]$ be a subset of the universal variables.

    Then $\abs(\Psi_1,V')\equiv \abs(\Psi_2,V')$ implies $\Psi_1\equiv\Psi_2$.
\end{lemma}

\begin{proof}
    For the sake of simplicity, we assume that $V'$ contains a single
    universal variable, \ie $V' = \{x\}$ for some $x\in\varall[\prefix]$. All other cases
    for $V'$ can be obtained by applying abstraction multiple times, because
    $\abs\bigl(\abs(\Psi,V_1), V_2\bigr) = \abs(\Psi,V_1\dcup V_2)$ for all
    disjoint sets $V_1,V_2\subseteq \varall$.

    So let $V' \colonequals \{x\}$ for some $x\in\varall[\prefix]$. Assume that the claim was wrong,
    \ie $\abs(\Psi_1,V')\equiv \abs(\Psi_2,V')$, but $\Psi_1\not\equiv\Psi_2$. \Wlogen
    let $(s_y)_{y\in\varex[\prefix]}$ be functions which are a Skolem function
    for $\Psi_1$, but not for $\Psi_2$. Since $(s_y)_{y\in\varex[\prefix]}$ is not a
    Skolem function for $\Psi_2$, there must be an assignment $\lambda:\varall[\prefix]\to\bools$
    such that replacing all existential variables $y$ by $s_y$ and then all
    universal variables $x$ by $\lambda(x)$ makes $\varphi_1$ true and $\varphi_2$ false.

    Now we use $\lambda$ and $(s_y)_{y\in\varex[\prefix]}$ to construct new functions
    $\lambda'$ and $(s'_z)_{z\in\varex[\prefix]\cup V'}$. First, $\lambda'$ is simply the
    restriction of $\lambda$ to $\varall[\prefix]\setminus V'$. Second, we set
    \[
        s'_v \colonequals
        \begin{cases}
            \lambda(v) & \text{for $v\in V'$,} \\
            {s_v}_{|\lambda} & \text{for $v\in\varex[\prefix]$.}
        \end{cases}
    \]
    Applying first $(s'_z)_{z\in\varex[\prefix]\cup V'}$ and then $\lambda'$ to
    $\varphi_1$ and $\varphi_2$ has the same effect as applying first
    $(s_z)_{z\in\varex[\prefix]}$ and then $\lambda$ to the same formulas. That means
    $(s'_z)_{z\in\varex[\prefix]\cup V'}$ are Skolem functions for $\abs(\Psi_1,V')$,
    but not for $\abs(\Psi_2,V')$. Therefore $\abs(\Psi_1,V')\not\equiv \abs(\Psi_2,V')$.
    This is a contradiction to our assumption.\qed
\end{proof}

Note that the contrary does not hold: If $\Psi_1\equiv\Psi_2$, then
$\abs(\Psi_1,V')$ is not necessarily equivalent to $\abs(\Psi_2,V')$.
\begin{example}
For instance consider the formulas
\[
    \Psi_1 \colonequals \forall x\exists y(\emptyset): (x\land y)\qquad\text{and}\qquad
    \Psi_2 \colonequals \forall x\exists y(\emptyset): (x\land y)\land\neg x.
\]
Both DQBFs are unsatisfiable, \ie they have the same Skolem functions (none) and
are therefore equivalent. However,
$\abs(\Psi_1,\{x\}) = \exists x(\emptyset)\exists y(\emptyset): x\land y$ is satisfiable with $s_x = 1$ and $s_y = 1$.
$\abs(\Psi_2,\{x\}) = \exists x(\emptyset)\exists y(\emptyset): (x\land y)\land\neg x$ is unsatisfiable.
\end{example}

\begin{definition}[Clause Derivation]
    \label{def:clause_derivation}
    Let $\Psi\colonequals\prefix:\varphi$ be a DQBF and $C$ a compatible clause.
    If $\prefix:\varphi\land\neg C\upy\conflict$ and
    $\prefix:\varphi\ \ \vDash\ \ \prefix:\varphi\land C$,
    then we write $\Psi\upd C$ to denote that $C$ can be derived from $\Psi$
    by unit propagation.
\end{definition}

One might wonder why the additional requirement $\prefix:\varphi\vDash\prefix:\varphi\land C$
is necessary. In case of a SAT problem, it is not. If unit propagation applied to
$\varphi\land\neg C$ (without universal reduction) yields a conflict, then
$\varphi$ and $\varphi\land C$ are equivalent.
However, for (D)QBF, the fact that unit propagation (with universal reduction) leads to
a conflict is not sufficient to make the addition of $C$ sound:
\begin{example}[Adapted from \cite{LonsingE18}]
    Consider the DQBF $\Psi$ with
    \[
        \Psi\colonequals \prefix:\varphi\colonequals\forall x_1\exists y_1(x_1):
            (x_1\lor\neg y_1)\land (\neg x_1\lor y_1)\,.
    \]
    and clause $C = y_1$. We have
    $\prefix:\varphi\land\neg C\upy\conflict$ since propagating $\neg C= \neg y_1$
    produces $\neg x_1$ due to the second clause of $\Psi$. Universal reduction reduces
    it to the empty clause. However, $\prefix:\varphi\nvDash\prefix:\varphi\land C$ since
    $\prefix:\varphi$ is satisfiable with Skolem function $s_{y_1}(x_1) = x_1$, while
    $\prefix:\varphi\land C$ is unsatisfiable.
\end{example}
\begin{lemma}
    \label{lemma:up_clause}
    Let $\Psi\colonequals \prefix:\varphi$ be a DQBF and $C$
    a compatible clause such that $\dep(C)=\emptyset$.

    If $(\prefix:\varphi\land\neg C)\upy\conflict$, then
    $\prefix:\varphi\ \ \equiv\ \ \prefix:\varphi\land C$.
\end{lemma}

\begin{proof}
    We set $\Psi'\colonequals\prefix:\varphi\land C$.
    Assume that the claim is wrong, \ie that there is a function $(s_y)_{y\in\varex}$,
    which is a Skolem function for $\Psi$, but not for $\Psi'$. So all literals in
    $C$ are mapped to $0$ by $(s_y)_{y\in\varex}$. Note that $\dep(C)=\emptyset$ implies
    that $C$ only contains existential variables whose Skolem functions are constant.

    On the other hand, since $\prefix:\varphi\land\neg C\upy\conflict$,
    the DQBF $\prefix:\varphi\land\neg C$ is unsatisfiable. Because $(s_y)_{y\in\varex}$ is a Skolem
    function for $\prefix:\varphi$, turning all clauses of $\varphi$ into tautologies,
    only $\neg C$ cannot be satisfied by $(s_y)_{y\in\varex}$. That means at least
    one literal $\neg\ell$ of $\neg C$ is mapped to $0$.
    This, however, implies that $C$ is satisfied by $(s_y)_{y\in\varex}$, because
    $\ell\in C$ is mapped to $1$.

    Consequently, we have a contradiction, and our assumption must be wrong.\qed
\end{proof}

\begin{lemma}
    \label{lemma:up_abs}
    Let $\Psi \colonequals\prefix:\varphi$ be a DQBF, $C$ a compatible clause, and $V' \colonequals\dep(C)$.
    If $\abs(\prefix:\varphi\land\neg C,V')\upy\conflict$, then
    $\abs(\prefix:\varphi, V')\ \ \equiv\ \ \abs(\prefix:\varphi\land C, V')$.
\end{lemma}
\begin{proof}
    The claim follows from \cref{lemma:up_clause} since all variables that
    appear in $C$ are existentially quantified in $\abs(\prefix:\varphi\land\neg C, V')$ and
    have empty dependency sets.\qed
\end{proof}

\begin{theorem}
    \label{th:add_clause}
    Let $\Psi \colonequals\prefix:\varphi$ be a DQBF, $C$ a compatible clause,
    and $V'\colonequals\dep(C)$.
    If $\abs(\prefix:\varphi\land\neg C,V')\upy\conflict$, then
    $\prefix:\varphi\ \ \equiv\ \ \prefix:\varphi\land C$.
\end{theorem}

\begin{proof}
    This theorem immediately follows from \cref{lemma:up_abs,lemma:abs_equiv}.\qed
\end{proof}

\subsection{Vivification}
\label{sec:vivi}

\noindent\Cref{th:add_clause} provides the theoretical
foundation for generalizing vivification~\cite{PietteHS08}. The original
version was defined for quantifier-free formulas and therefore
uses unit propagation without universal reduction.

\begin{theorem}[Vivification]
    \label{th:vivification}
    Let $\Psi \colonequals \prefix:\varphi\land C$ be a DQBF with a clause $C$, $C'\subsetneq C$, and
    $V' \colonequals \dep(C')$.
    \begin{itemize}
    \item If $\abs(\prefix:\varphi\land C',V')\upy\conflict$, then
        $\Psi\ \equiv\ \prefix:\varphi\land C'$.

    \item If $\abs(\prefix:\varphi\land C', V')\upy U$ with $\ell\in U\cap (C\setminus C')$,
        then $\Psi\ \equiv\ \prefix:\varphi\land (C'\cup\{\ell\})$.
    \end{itemize}
\end{theorem}
\begin{proof}
    We prove the two statements of the theorem:
    \begin{itemize}
        \item If $\abs(\prefix:\varphi\land \neg C', V')\upy\conflict$,
            $\Psi$ is equivalent to $\prefix:\varphi\land C'$ according to
            \cref{th:add_clause}.
        \item Assume $\abs(\prefix:\varphi\land\neg C', V')\upy U$
            with $\ell\in U\cap (C\setminus C')$. Since unit propagation
            applied to $\abs(\prefix:\varphi\land\neg C', V')$ yields the
            unit literal $\ell$, applying unit propagation to
            $\abs(\prefix:\varphi\land\neg C'\land\neg\ell, V')$ leads to
            a conflict. \Cref{th:add_clause} tells us that $\Psi$ is
            equivalent to $\prefix:\varphi\land (C'\cup\{\ell\})$.
    \end{itemize}\qed
\end{proof}

\subsection{Unit Propagation Lookahead (UPLA)}
\label{sec:upla}

\begin{lemma}
    \label{lemma:upla}
    Let $\Psi\colonequals\prefix:\varphi$ be a DQBF and $\ell$ a literal.
    We set $V'\colonequals\dep(\ell)$.

    If $\abs(\prefix:\varphi\land\ell, V')\upy U$ with $\kappa\in U$, then
    $\abs(\prefix:\varphi, V')\ \equiv\ \abs(\prefix:\varphi\land (\neg\ell\lor\kappa), V')$.
\end{lemma}

\begin{proof}
    It is clear that every Skolem function for $\abs(\prefix:\varphi\land(\neg\ell\lor\kappa),V')$
    is a Skolem function for $\abs(\prefix:\varphi,V')$ as well.

    So let $(s_y)_{y\in\varex\cup V'}$ be a Skolem function for $\abs(\prefix:\varphi,V')$.
    Since $\var(\ell)$ is existential in the abstraction and has an empty dependency set,
    $s_{\var(\ell)}$ is either constantly $0$ or constantly $1$. If $s_{\var(\ell)}$ is such
    that $\ell$ is $0$, $(s_y)_{y\in\varex\cup V'}$ satisfies $(\neg\ell\vee\kappa)$
    and is therefore a Skolem function for $\abs(\prefix:\varphi\land(\neg\ell\vee\kappa))$.

    If $\ell$ is constantly $1$, then $(s_y)_{y\in\varex\cup V'}$ is a Skolem function for
    $\abs(\prefix:\varphi\land\ell, V')$. Since $\abs(\prefix:\varphi\land\ell, V')\upy U$,
    \[
        \abs\bigl(\prefix:\varphi\land\ell, V'\bigr)\ \equiv\
        \abs\bigl(\prefix:\varphi\land\ell\land\bigwedge_{m\in U}m, V'\bigr)
    \]
    according to \cref{lemma:unit_propagation}.
    Therefore, all Skolem functions of $\abs(\prefix:\varphi\land\ell)$ have to make
    all literals in $U$ true, in particular $\kappa\in U$. Therefore
    $(s_y)_{y\in\varex\cup V'}$ satisfies $\kappa$ and also $(\neg\ell\lor\kappa)$.
    \qed
\end{proof}

\begin{theorem}
    \label{th:upla}
    Let $\Psi\colonequals\prefix:\varphi$ be a DQBF and $v$ a variable. We set
    $V'\colonequals\dep(v)$.

    Let $U^0$ and $U^1$ be such that $\abs(\prefix:\varphi\land v, V')\upy U^1$ and
    $\abs(\prefix:\varphi\land\neg v, V')\upy U^0$, provided that no conflict occurs;
    otherwise we set $U^1\colonequals\emptyset$ ($U^0\colonequals\emptyset$).
    \begin{itemize}
    \item If $\abs(\prefix:\varphi\land v, V')\upy\conflict$, then
        $\Psi\ \equiv\ \prefix:\varphi\land\neg v$; \\
        if $\abs(\prefix:\varphi\land\neg v, V')\upy\conflict$, then
        $\Psi\ \equiv\ \prefix:\varphi\land v$.

    \item $\Psi\ \equiv\prefix:\varphi\land
        \bigwedge\limits_{\kappa\in U^0\cap U^1}\kappa.$

    \item $\Psi\ \equiv\prefix:\varphi\land
        \bigwedge\limits_{\kappa: \kappa\in U^1\land\neg\kappa\in U^0}(\ell\equiv\kappa).$
    \end{itemize}
\end{theorem}
\begin{proof}
    We prove the three statements of this theorem:
    \begin{itemize}
    \item The first statement is a direct consequence of \cref{th:add_clause}.

    \item For the second statement, assume that $\kappa\in U^1\cap U^0$. According to
    \cref{lemma:upla}, we have
    \[
    \abs(\prefix:\varphi,V')
    \quad\equiv\quad\abs(\prefix:\varphi\land(\neg v\vee\kappa), V')
    \quad\equiv\quad\abs(\prefix:\varphi\land(v \vee\kappa), V').
    \]
    Let $(s_y)_{y\in\varex\cup V'}$ be a Skolem function of these formulas.
    We need to show that it makes $\kappa$ constantly true. Since $v$ is a
    existential variable with empty dependencies in the abstraction, the Skolem
    function $s_v$ is either constantly $1$ or $0$.

    If $s_v$ is constantly $1$, the formula $\abs(\prefix:\varphi\land(\neg v\vee\kappa),V')$
    requires that the Skolem function makes $\kappa$ constantly $1$. In case that
    $s_v$ is constantly $0$, the formula $\abs(\prefix:\varphi\land(\ell\vee\kappa),V')$
    requires the same. So we can conclude that every Skolem function for $\Psi$ makes
    $\kappa$ constantly $1$. Consequently,
    $\abs(\prefix:\varphi,V')\ \equiv\ \abs(\prefix:\varphi\land\kappa, V')$.
    Now, \cref{lemma:abs_equiv} implies that $\Psi\equiv\prefix:\varphi\land\kappa$.

    \item Let $\kappa$ be a literal with $\kappa\in U^1$ and $\neg\kappa\in U^0$.
    According to \cref{lemma:upla}, we have
    \[
        \abs(\prefix:\varphi,V')
        \quad\equiv\quad\abs(\prefix:\varphi\land(\neg v\vee\kappa), V')
        \quad\equiv\quad\abs(\prefix:\varphi\land(v\vee\neg\kappa), V').
    \]
    Let $(s_y)_{y\in\varex\cup V'}$ be a Skolem function of these formulas.
    We need to show that $v$ and $\kappa$ always have the same value.
    Since $v$ is an existential variable with empty dependencies in
    the abstraction, the Skolem function $s_v$ is either constantly $1$ or $0$.

    If $s_v = 1$, we can conclude in a similar way as
    in the proof of the second statement that $\kappa$ need to be constantly
    $1$ as well. Otherwise, if $s_v = 0$, we obtain in
    the same way that $\kappa$ needs to be constantly $0$ as well.
    That means the Skolem functions of $v$ and $\kappa$ are the same.
    Therefore they satisfy
    $(v\equiv\kappa)$, so $\abs(\prefix:\varphi,V')\ \equiv\
    \abs(\prefix:\varphi\land(v\equiv\kappa), V')$.
    \Cref{lemma:abs_equiv} implies that $\Psi\ \equiv\ \prefix:\varphi\land(v\equiv\kappa)$.\qed
    \end{itemize}
\end{proof}

\subsection{Identifying Redundant Clauses}
\label{sec:clause_redundancy}

\noindent In this section, we discuss techniques based on unit propagation which allow
to determine that certain clauses are redundant. They can be deleted from the formula
without changing satisfiability of DQBFs.

\subsubsection{Asymmetric Tautologies}
\label{ssec:dqat}

\begin{definition}[DQAT]
    Let $\Psi\colonequals\prefix:\varphi$ be a DQBF, $C$ a compatible clause, and $V'\colonequals\dep(C)$.
    Clause $C$ has property DQAT (dependency quantified asymmetric tautology) \wrt $\Psi$
    if $\abs(\prefix:\varphi\land\neg C,V')\upy\conflict$.
\end{definition}

\begin{theorem}
    If $C$ has property DQAT \wrt a DQBF $Q:\varphi$, then
    $\prefix:\varphi$ and $\prefix:\varphi\land C$ are equivalent.
\end{theorem}
This theorem directly follows from \cref{th:add_clause}.

We can conclude that a clause $C$ is redundant, if unit propagation applied
to $\abs\bigl(\prefix:\varphi\land\neg C, \dep[\prefix](C)\bigr)$ yields a conflict.

\subsubsection{Resolution Asymmetric Tautologies}

\noindent Resolution asymmetric tautologies (RAT) are an important proof system for refuting
Boolean satisfiability (SAT). Determining whether a clause has the RAT property
depends on the resolvents \wrt a pivot literal from the clause. In case of quantified
formulas, we need to distinguish whether this literal is existential or universal.

\begin{definition}[Outer Variables, \cite{Blinkhorn20}]
    Let $\Psi\colonequals\prefix:\varphi$ be a DQBF over variables $V$ and $v\in V$. \\
    If $v\in\varex$ is an existential variable, we set
    \[
    \OV(\prefix,v) \colonequals
        \bigl\{
            w\in V\,\big|\,\dep(w)\subseteq\dep(v)
        \bigr\}\,.
    \]
    If $v\in\varall$ a universal variable, we set
    \begin{align*}
        S_v &\colonequals \{ y\in\varex\,|\, v\in D_y \} &&\text{the variables dependent on $v$},\\
        I_v &\colonequals \{ y\in\varex\,|\, v\notin D_y\} && \text{the variables independent from $v$},\\
        K_v &\colonequals \bigcap\limits_{y\in S_v} D_y && \text{the \emph{kernel} of $v$,}\\
        \OV(\prefix,v) &\colonequals K_v\cup \{ y\in I_v \,|\, D_y\subseteq K_v \}
                &&\text{the outer variables of $v$.}
    \end{align*}
\end{definition}

\begin{definition}[Outer Clause, Outer Resolvent]
    The \emph{outer clause} of a clause $C$ on a literal $\ell\in C$ \wrt prefix $\prefix$
    is the clause
    \[
        \OC(\prefix,C,\ell) \colonequals \bigl\{ \kappa\in C\,\big|\, \var(\kappa)\in\OV(\prefix,\var(\ell))\bigr\}\,.
    \]
    Let $D\in\varphi$ be a clause with $\neg\ell\in D$. The \emph{outer resolvent} of
    $C$ and $D$ \wrt $\ell$ is given by
    \[
        \OR(\prefix,C,D,\ell)\colonequals
        \begin{cases}
            C\cup\bigl(\OC(\prefix,D,\neg\ell)\setminus\{\neg\ell\}\bigr)& \text{if $\ell$ is existential,}\\
            \bigl(C\setminus\{\ell\}\bigr)\cup\OC(\prefix,D,\ell) & \text{if $\ell$ is universal.}
        \end{cases}
    \]
\end{definition}

\begin{definition}[\dqior, \cite{Blinkhorn20}]
    Let $\Psi\colonequals\prefix:\varphi$ be a DQBF and $C$ a clause. $C$ has property
    \dqior{} (dependency quantified implied outer resolvent) on a literal $\ell\in C$
    when
    \[
        \Psi\vDash\OR(\prefix,C,D,\ell)
            \qquad\text{for all $D\in\varphi$ with $\neg\ell\in D$.}
    \]
\end{definition}

\begin{theorem}[\cite{Blinkhorn20}]
    Let $\prefix:\varphi$ be a DQBF and $C$ a compatible clause.
    \begin{itemize}
        \item If $C$ has property
            \dqior{} on an existential literal $\ell\in C$, then
            \[
                \prefix:\varphi\quad\text{and}\quad
                \prefix:\varphi\land C
            \]
            are equi-satisfiable.
        \item If $C$ has property \dqior{} on a universal literal $\ell\in C$, then
            \[
                \prefix: \varphi\land C\quad\text{and}\quad
                \prefix: \varphi\land (C\setminus\{\ell\})
            \]
            are equi-satisfiable.
        \end{itemize}
\end{theorem}

When we replace implication ($\vDash$) in \dqior{} by unit propagation, we arrive
at a property named \emph{dependency quantified resolution asymmetric tautology}
(DQRAT). The original version for DQBF by \cite{Blinkhorn20} uses unit propagation
without universal reduction. Here we prove a stronger result by allowing
universal reduction during unit propagation, combining results from
\cite{Blinkhorn20} and \cite{LonsingE18}.

\begin{definition}[\dqratPlus]
    A compatible clause $C$ has property \dqratPlus{} on a literal $\ell$ \wrt a DQBF
    $\Psi \colonequals \prefix:\varphi$ iff
    \[
        \abs(\prefix:\varphi\land E, V')\upy\conflict
    \]
    for all clauses $D\in\varphi$ with $\neg\ell\in D$, where
    $E\colonequals\OR(\prefix,C,D,\ell)$ and $V' \colonequals \dep(E)$.
\end{definition}

For more details on DQRAT (without applying universal reduction during unit propagation),
see also \cite{Blinkhorn20}.

\begin{theorem}
    \label{th:dqrat_plus}
    Let $C$ be a compatible clause with property \dqratPlus{} \wrt a DQBF $\prefix:\varphi$ and
    $\ell\in C$.
    \begin{itemize}
        \item If $\ell$ is existential, then
            $\prefix:\varphi$ and $\prefix:\varphi\land C$ are equi-satisfiable.
        \item If $\ell$ is universal, then
            $\prefix:\varphi\land C$ and $\prefix:\varphi\land (C\setminus\{\ell\})$
            are equi-satisfiable.
    \end{itemize}
\end{theorem}

\begin{proof}
    It is sufficient to show that \dqratPlus{} implies \dqior{}.

    Let $D$ be a clause with $\neg\ell\in D$ and
    $E\colonequals\OR(\prefix,C,D,\ell)$. We set $V' \colonequals \dep(E)$.
    If $\abs(\prefix:\varphi\land\neg E, V')\upy\conflict$,
    \cref{th:add_clause} implies that $\Psi$ is equivalent
    to $\prefix:\varphi\land E$. That means in particular,
    $\Psi\vDash E$.
    If $C$ has property DQIOR, this holds for all such clauses
    $D\in\varphi$ with $\neg\ell\in D$; therefore
    \dqratPlus{} implies DQIOR.\qed
\end{proof}

\section{Conclusion}

\noindent In this paper, we have closed a gap in the theory of preprocessing techniques
for DQBF by showing how unit propagation can be combined with universal reduction in the
application of vivification, unit propagation look-ahead, and the identification of
clauses as dependency quantified resolution asymmetric tautologies. These results generalize
similar results for QBF by Losing and Egly~\cite{LonsingE18,LonsingE19,Lonsing19} from
QBF to DQBF and also extend the definition of DQRAT by Blinkhorn~\cite{Blinkhorn20}.

As future work, we plan to extend the DQBF preprocessor HQSpre~\cite{wimmer-et-al-jsat-2019}
by the improved techniques and to evaluate their effectiveness.

\section*{References}
\bibliographystyle{plainurl}
\bibliography{univ_reduction}

\end{document}